# Gapless dual-comb spectroscopy in terahertz region

Takeshi Yasui[1, 2,*], Yi-Da Hsieh[2], Yuki Iyonaga[2], Yoshiyuki Sakaguchi[2], Shuko Yokoyama[2, 3], Hajime Inaba[4], Kaoru Minoshima[4], Francis Hindle[5], and Tsutomu Araki[2]

[1]Institute of Technology and Science, The University of Tokushima, Tokushima, Tokushima 770-8506, Japan
[2]Graduate School of Engineering Science, Osaka University, Toyonaka, Osaka 560-8531, Japan
[3]Micro Optics Co.,Ltd, Nishikyo, Kyoto, Kyoto 610-1104, Japan
[4]National Metrology Institute of Japan, National Institute of Advanced Industrial Science and Technology, Tsukuba, Ibaraki 305-8563, Japan
[5]Laboratoire de Physico-Chimie de l'Atmosphère, Université du Littoral Côte d'Opale, Dunkerque 59140, France

Corresponding author: yasui.takeshi@tokushima-u.ac.jp

Abstract

We demonstrated combination of gapless terahertz (THz) comb with dual-comb spectroscopy, namely gapless dual-THz-comb spectroscopy, to achieve the spectral resolution equal to width of the THz comb tooth. The gapless THz comb was realized by interpolating frequency gaps between the comb teeth with sweeping of a laser mode-locked frequency. The demonstration of low-pressure gas spectroscopy with gapless dual-THz-comb spectroscopy clearly indicated that the spectral resolution was decreased down to 2.5-MHz width of the comb tooth and the spectral accuracy was enhanced to $10^{-6}$ within the spectral range of 1THz. The proposed method will be a powerful tool to simultaneously achieve high resolution, high accuracy, and broad spectral coverage in THz spectroscopy.

# 1. Introduction

Optical frequency comb has emerged as a new mode for optical frequency metrology and broadband spectroscopy because its teeth can be used as a precise ruler in the frequency domain [1]. Since the first demonstration of near-infrared (NIR) optical comb [2], the wavelength of the optical comb has been successively extended from extreme ultraviolet (XUV) region to terahertz (THz) region in combination with the nonlinear wavelength conversion [3-5]. In the beginning, a single tooth of the optical comb has been selected and used to determine the absolute value of optical frequency for the metrology. On the other hand, a series of the comb teeth can be used as frequency markers for the broadband spectrum. In particular, its interesting feature for broadband spectroscopy is in ultra-wide dynamic range of frequency scale, defining as a ratio of full spectral range of the comb to the spectral resolution. To use such the feature for spectroscopy, the optical comb has to be combined with Michelson-based Fourier-transform spectrometer (FTS) [6]. Unfortunately, the spectral resolution in usual FTS is insufficient to spectrally resolve each tooth in the optical comb because of too densely distributed teeth.

Recently, dual-comb spectroscopy has opened a new door for the optical-comb-based FTS because it enables us to make full use of all teeth in the optical comb [4]. In this method, one optical comb for measurement (frequency interval = mode-locked frequency = $f_1$) is interacted with another optical comb with slight frequency offset (frequency interval = $f_2$ = $f_1$ + $\Delta f$) acting as a local oscillator. Multi-frequency-heterodyning process between the two combs results in the generation of a secondary frequency comb in the radio-frequency (RF) region, namely RF comb (frequency interval = $\Delta f$). Since the RF comb is a replica of the measurement comb only downscaled by a ratio of $f_1/\Delta f$ in frequency, one can utilize all the teeth in the optical comb easily via direct observation of the RF comb and calibration of the frequency scale. Since the dual-comb spectroscopy features high sensitivity, rapid data acquisition, and measurement of optical electric filed as well as high spectral resolution and accuracy, it has rapidly spread to other wavelength regions [5, 7-9]. However, even in these cases of dual-comb spectroscopy, it is still difficult to make full use of the ultra-wide dynamic range of frequency scale inherent in optical comb. This is because the practical spectral resolution should be equal to its frequency spacing between the comb teeth due to discrete distribution of them. On the other hand, focusing the fact that the linewidth of each tooth is much narrower than their frequency spacing, the spectral resolution and accuracy will be largely enhanced if this narrow linewidth can be used for the spectroscopy in place of the frequency spacing. To this end, frequency gap between the comb teeth should be interpolated.

In this article, we propose a combination of a gapless THz comb with dual-comb spectroscopy, namely gapless dual-THz-comb spectroscopy, by precise sweeping of frequency interval in dual optical combs. Furthermore, the gapless dual-THz-comb spectroscopy is applied for high-precision spectroscopy of gas-phase molecules in low pressure.

## 2. Principle

First, let us consider how we can acquire the hyperfine-structured spectrum of a THz comb, which surpasses the spectral resolution in FTS. Although the RF comb can be directly acquired with an RF spectrum analyzer based on multi-frequency-heterodyning photoconductive detection [5], this frequency-domain measurement was hampered by low efficiency of signal acquisition [10]. Here, we focus on data acquisition in time domain. A THz time-domain spectroscopy (THz-TDS), which is a typical FTS using the pulsed THz electric field, measures the temporal waveform of only a single THz pulse with the mechanical time-delay scanning, as shown in Fig. 1(a). Taking the Fourier transformation (FT) of the waveform gives the continuous spectrum of the broadband THz radiation without comb structure, as shown in Fig. 1(b). On the other hand, if the time window is greatly expanded to cover more than one pulse period, one can measure the temporal waveform of consecutive THz pulses, that is to say, a THz pulse train, as shown in Fig. 1(c). Taking the FT of this will give the hyperfine-structured spectrum of the THz comb because FT of periodical THz pulses imprints frequency modulation on broadband THz spectrum, as shown in Fig. 1(d).

However, a conventional THz-TDS system is not practical for measuring such the temporal waveform of THz pulse train because it requires scanning with quite a long time delay. Instead, therefore, we focus on asynchronous-optical-sampling THz-TDS (ASOPS-THz-TDS) using two mode-locked lasers with slightly mismatched mode-locked frequencies [10-13]. Since ASOPS-THz-TDS enable us to expand time scale of ps THz pulse up to µs order accurately based on the principle of asynchronous optical sampling [14], the resulting slow temporal waveform can be measured directly by a standard oscilloscope without the need for the mechanical time-delay scanning. Therefore, one can arbitrarily adjust the observed time window of THz pulse temporal waveform by changing the time scale of the oscilloscope. This allows the temporal waveform of the THz pulse train to be measured [see Fig. 1(c)] [15].

Next, we consider how we can interpolate frequency gap between the comb teeth. THz comb is a harmonic frequency comb of a laser mode-locked frequency without carrier-envelope-offset frequency, and its frequency spacing is exactly equal to the mode-locked frequency. Therefore, the absolute frequency of each comb teeth can be tuned by changing the mode-locked frequency. If incremental sweeping of the comb tooth is repeated at an interval equal to the tooth linewidth and the resulting all of the comb spectra are overlaid in the spectral domain as shown in Fig. 1(e), the frequency gaps between the comb teeth can be fully interpolated. In this way, the gapless THz comb can be achieved. This situation is equivalent to continuous sweeping of a single-mode, narrow-linewidth CW-THz wave across a broadband THz spectral region. The resulting spectral resolution should be equal to the linewidth of the comb tooth. Therefore, the combination of the gapless THz comb with the dual-comb spectroscopy enables us to make full use of ultra-wide dynamic range of frequency scale in the THz comb.

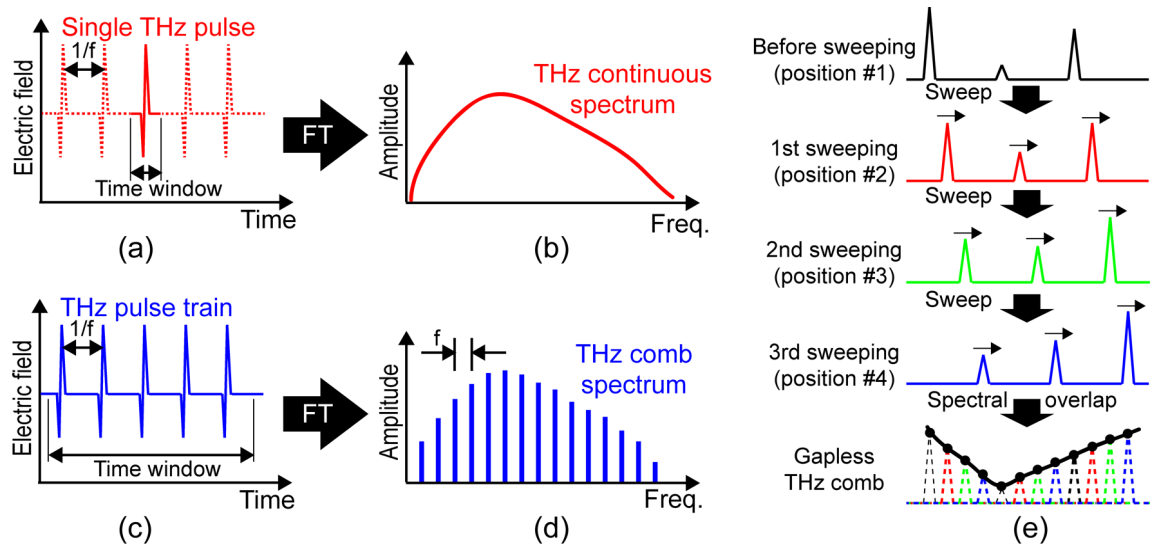

Fig. 1. (a) Temporal waveform of single THz pulse and (b) corresponding THz continuous spectrum. (c) Temporal waveform of THz pulse train and (d) corresponding THz comb spectrum. (e) Gapless THz comb achieved by incremental sweeping of THz comb tooth and spectral overlapping of them.

### 3. Experimental setup

Figure 2 illustrates a schematic diagram of our ASOPS-THz-TDS system, consisting of dual mode-locked Er-fiber lasers (ASOPS TWIN 250 with P250, Menlo Systems; center wavelength $\lambda_c$ = 1550 nm, pulse duration $\Delta\tau$ = 50 fs, mean power $P_{mean}$ = 500 mW) and a delay-stage-free THz-TDS setup including a pair of photoconductive antenna (PCA) [15]. The individual mode-locked frequencies of the two lasers ($f_1$ = 250,000,000 Hz and $f_2$ = 250,000,050 Hz) and the frequency offset between them ($\Delta f = f_2 - f_1$ = 50 Hz) were stabilized by two independent laser control systems referenced to a rubidium frequency standard (Rb-FS, accuracy = $5\times10^{-11}$ and instability = $2\times10^{-11}$ at 1 s). Furthermore, $f_1$ and $f_2$ can be respectively tuned within the frequency range of ±0.8 % by changing reference frequencies synthesized from the frequency standard. After wavelength conversion of the two laser beams by second-harmonic-generation crystals (SHGs), pulsed THz radiation was emitted by a dipole-shaped, low-temperature-grown (LTG), GaAs PCA (PCA1) triggered by pump light ($\lambda_c$ = 775 nm, $\Delta\tau$ = 80 fs, $P_{mean}$ = 19 mW), and was then detected by another dipole-shaped LT-GaAs-PCA (PCA2) triggered by probe light ($\lambda_c$ = 775 nm, $\Delta\tau$ = 80 fs, $P_{mean}$ = 9 mW). Portions of the output light from the two lasers were fed into a sum-frequency-generation cross-correlator (SFG-XC). The resulting SFG signal was used to generate a time origin signal in the ASOPS-THz-TDS. After amplification with a current preamplifier (AMP, bandwidth = 1MHz, gain = 4 ×10$^6$ V/A), the temporal waveform of the output signal from PCA2 was acquired with a digitizer (sampling rate = 2 MS/s, resolution = 20 bit) by using the SFG-XC's output as a trigger signal and the frequency standard's output as a clock signal, respectively. Then, time scale of the observed signal was calibrated by a temporal magnification factor of $f_1/\Delta f$ (= 250,000,000/50 = 5,000,000) [12]. The sampling rate and the temporal magnification factor enable us to measure the temporal waveform of the pulsed THz electric field at a sampling interval of 100 fs. Finally, the amplitude spectrum of THz comb was obtained by FT of the temporal waveform of THz pulse train. To suppress pressure broadening of absorption lines, molecular gasses were enclosed in a low-pressure gas cell (length = 500 mm, diameter = 40 mm). The optical path, in which the THz beam propagated, except for the gas cell, was purged with dry nitrogen gas to avoid absorption by atmospheric moisture.

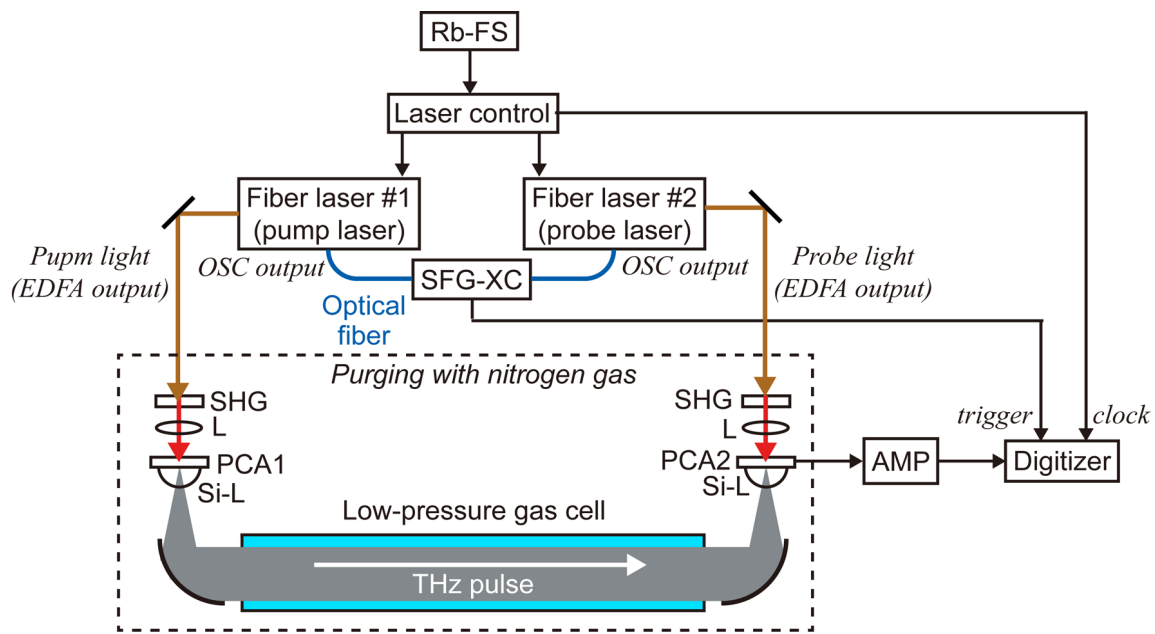

Fig. 2. Experimental setup. Rb-FS, rubidium frequency standard; SFG-XC, sum-frequency-generation cross-correlator; SHGs, second-harmonic-generation crystals; Ls, lenses; EDFA, erbium-doped fiber amplifier; OSC, erbium-doped fiber oscillator; PCA1, dipole-shaped LT-GaAs photoconductive antenna for THz generation; PCA2, dipole-shaped LT-GaAs photoconductive antenna for THz detector; Si-Ls, silicon lenses; AMP, current preamplifier.

## 4. Results

### 4.1 Spectroscopy of low-pressure water vapor

To evaluate the spectral resolution in the proposed method, we measured the rotational transition $1_{10} \leftarrow 1_{01}$ at 0.557 THz in water vapor. To reduce the pressure

-5-

broadening and the strong absorption, we diluted the water vapor at 10 Pa with a nitrogen gas at 320 Pa. Under this pressure condition, the pressure broadening linewidth was expected to 23 MHz. Figure 3(a) shows the amplitude spectrum of the *gapped* THz comb expanded around 0.557 THz, obtained by taking the FFT of the temporal waveform of 10 consecutive THz pulses (time window = 40 ns, signal averaging = 5000, acquisition time = 1000 s). The comb teeth had a frequency spacing of 250 MHz and a linewidth of 25 MHz. The frequency spacing was equal to the laser mode-locked frequency, whereas the linewidth was consistent with the reciprocal of the temporal window. In this way, the amplitude spectrum without incremental sweeping of THz comb teeth did not indicate any spectral shape of the absorption line due to too coarse distribution of the comb teeth compared with the absorption linewidth of low-pressure water vapor.

We next demonstrated incremental sweeping of the comb teeth across the absorption line at 0.557 THz by changing the mode-locked frequencies of both lasers. Incremental increase of $f_1$ and $f_2$ by 11,210 Hz were repeated ten times while keeping $\Delta f$ at 50 Hz. The resulting overlaid spectra of comb modes are shown in Fig. 3(b). In this demonstration, a single shift of $f_1$ by 0.00448 % resulted in sweeping of the comb teeth by 10 % of its interval due to the large mode number of the THz comb (about 2,230 in this demonstration). Frequency gaps between comb teeth in Fig. 3(a) were fully interpolated. In other words, the gapless THz comb was successfully achieved. As a result, the sharp spectral dip was clearly appeared at the position of the water absorption line. Then, the absorption spectrum was obtained by extracting the peak amplitude of each comb teeth and normalization with a reference spectrum obtained under identical conditions, as shown in Fig. 3(c). The spectral linewidth was determined to be 24 MHz when a Lorentzian function was fitted to the spectral shape, indicated by the solid line in Fig. 3(c), which was consistent with the expected pressure broadening linewidth (= 23 MHz).

Next, we investigated pressure broadening of the same water line when the total pressure was changed. Black plots in Fig. 3(d) show the full-width at half-maximum (FWHM) of the observed absorption line with respect to the partial pressure of the water vapor, which was varied between 5 Pa and 160 Pa. For comparison, the theoretical curve is also indicated as a red line in Fig. 3(d) [16]. The experimental data were in good agreement with the theoretical curve down to 25 MHz, and then deviated from it. This result clearly indicated that the spectral resolution was enhanced down to the linewidth of the comb teeth (= 25 MHz) from their interval (= 250 MHz).

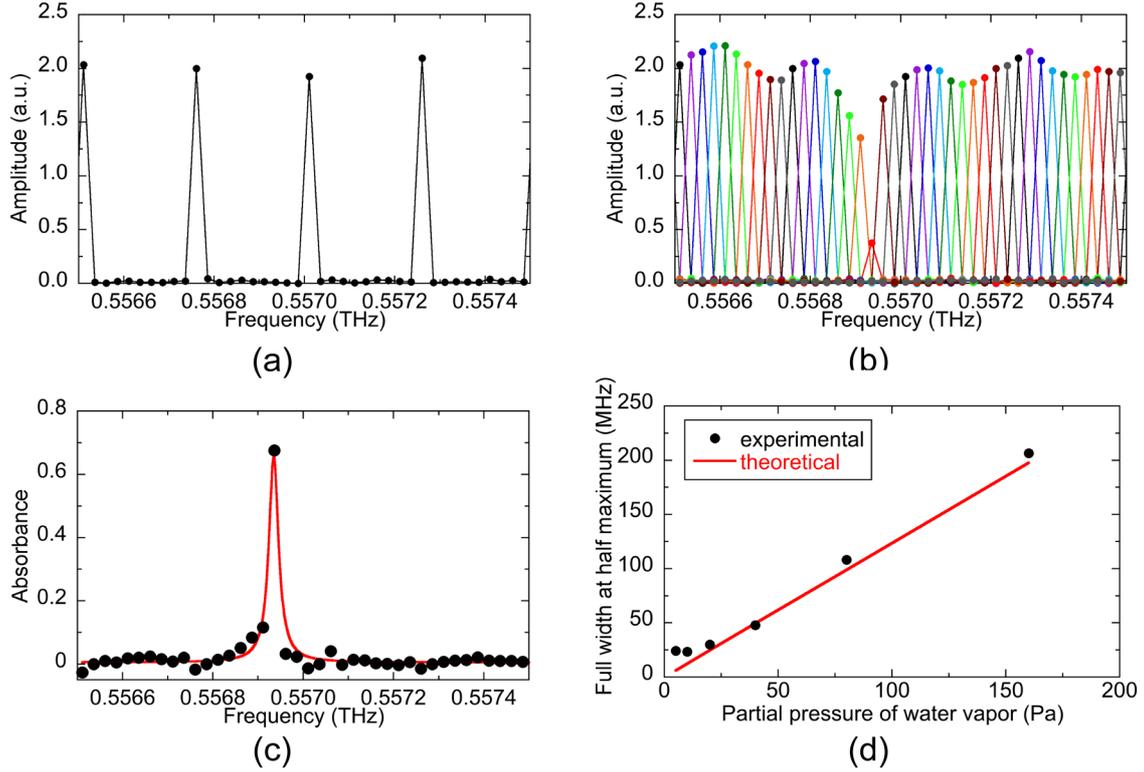

Fig. 3. Amplitude spectra of (a) gapped THz comb and (b) gapless THz comb around 0.557 THz after passing through low-pressure water vapor contained in the gas cell. (c) Absorption spectrum and (d) pressure broadening characteristics of the rotational transition $1_{10} \leftarrow 1_{01}$ in water vapor.

4.2 Spectroscopy of low-pressure acetonitrile gas

To demonstrate the capacity of the gapless dual-THz-comb spectroscopy to simultaneously probe multiple absorption lines of low-pressure molecular gas, we performed gas-phase spectroscopy of acetonitrile ($CH_3CN$). Since $CH_3CN$ is a symmetric top molecule with a rotational constant $B$ of 9.2 GHz, it indicates two features: manifold of rotational transitions regularly spaced by 2B (= 18.4 GHz) and hyperfine structure of rotational transitions determined by centrifugal distortion constant $D_{JK}$ [17]. Conventionally, these two features have been separately measured by using broadband THz-TDS [18] and a high-resolution CW-THz spectrometer [19]. Recently, one-pulse-period ASOPS-THz-TDS was successfully applied for observing both features of manifolds and the hyperfine structures simultaneously [13]. However, since its spectral resolution remained a laser mode-locked frequency (typically, several tens MHz to a few hundreds MHz), it is still difficult to fully resolve all the hyperfine structures. Here, we evaluated the possibility that the gapless dual-THz-comb spectroscopy allows us to observe the hyperfine structures more precisely due to the enhanced spectral resolution.

We first performed spectroscopy of a gas-phase $CH_3CN$ at 40 Pa enclosed in the gas cell with gapped THz comb, obtained by taking the FFT of the temporal waveform of 10 consecutive THz pulses (time window = 40 ns). Figure 4(a) shows the obtained absorption spectrum of this gas sample, in which a series of manifolds from J = 16 to J = 55 was clearly confirmed at interval of 18.4 GHz within a frequency range from 0.3 to 1.0 THz. We next applied a gapless THz comb with a tooth width of 25 MHz [see Fig. 3(b)] for high-precision spectroscopy, and observed the hyperfine structure in the manifold around 0.64 THz (J = 35 to J = 34), as shown in Fig. 4(b). To assign these absorption lines, we performed the multi-peak fitting analysis using a Lorentzian function, indicated by the red solid line in Fig. 4(b). From comparison with literature values reported in JPL database [20], we succeeded to assign lines K = 2 to 10 within a frequency discrepancy of 4.7 ± 2.6 MHz (mean ± standard deviation); however, we failed in assigning lines K = 0 and 1. This is because frequency separation of 12 MHz between them is too small to spectrally resolve it using the gapless THz comb with a tooth width of 25 MHz.

To spectrally resolve these two adjacent peaks, width and incremental step of comb teeth should be further reduced. To this end, we expanded the time window up to 400 ns, in which the temporal waveform of 100 consecutive THz pulses was measured. It should be emphasized that the time window of 400 ns is corresponding to the mechanical time-delay scanning by 60 m when traditional THz-TDS is used to make this experiment. Only ASOPS-THz-TDS can provide such the incomparable time window. This makes width of comb tooth to be reduced down to 2.5 MHz. When incremental sweeping of this comb tooth was repeated 15 times at intervals of 2.5 MHz across the spectral region of the two absorption lines of K = 0 and 1, two spectral dips were clearly appeared as shown in Fig. 4(c). We determined their center frequencies to be 0.643257578 THz for K = 0 and 0.643269502 THz for K = 1 by performing the multi-peak fitting analysis of the absorption spectrum, as shown in Fig. 4(d). Discrepancy of the center frequencies from those in JPL database [20] was 0.578 MHz for K = 0 and 0.502 MHz for K = 1, corresponding to the spectral accuracy of $10^{-6}$. However, it should be emphasized that this result dose not always indicate the limit of spectral accuracy in this method because literature values in JPL database includes an frequency error over 1 MHz. In other words, gapless dual-THz-comb spectroscopy may update the spectral database more precisely.

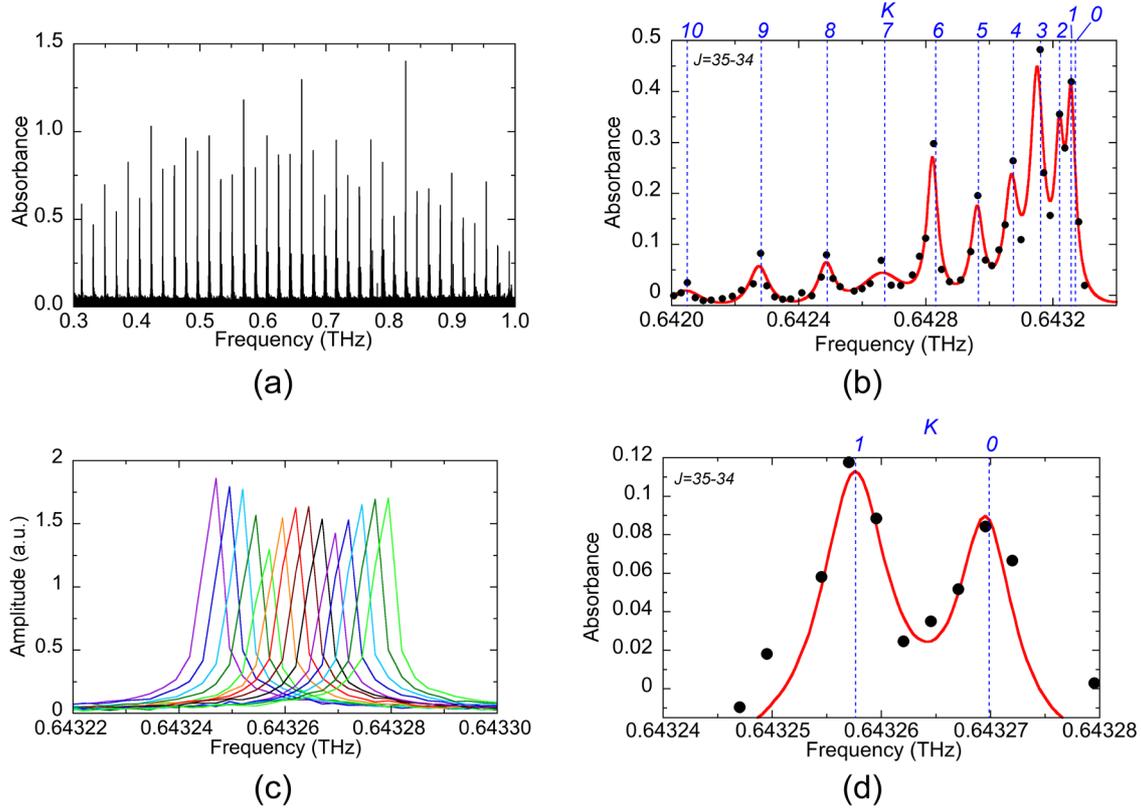

Fig. 4. Absorption spectrum of low-pressure acetonitrile gas (a) within a frequency range from 0.3 to 1 THz and (b) around 0.6428 THz. (c) amplitude spectrum and (d) absorption spectrum obtained by 15-times incremental sweeping of THz comb tooth across two adjacent absorption peaks.

## 6. Conclusion

To achieve the spectral resolution equal to width of the comb tooth, frequency gaps between THz comb teeth were successfully interpolated by simultaneous sweeping of the mode-locked frequencies in dual fiber combs. It is the first time to settle an inherent limitation common to all frequency combs including THz and optical combs, to the best of our knowledge. In combination with dual-comb spectroscopy, we demonstrated application of the gapless THz comb for high-precision spectroscopy of gas-phase molecules in low pressure. The resulting spectral resolution and accuracy achieved to 2.5 MHz and $10^{-6}$, respectively. Since the proposed combination of gapless comb with dual-comb spectroscopy is available in other frequency combs including optical comb, it will further enhance a performance of frequency-comb-based FTS.

## Acknowledgements

This work was supported by Collaborative Research Based on Industrial Demand from the Japan Science and Technology Agency, Japan. We also gratefully acknowledge financial support from the Renovation Center of Instruments for Science Education and Technology at Osaka University, Japan. The authors are grateful to Dr. M. Hashimoto of Osaka University, Japan, and Dr. H. Hoshina of RIKEN, Japan, for fruitful discussions.

## References

[1] Th. Udem, R. Holzwarth, and T. W. Hänsch, "Optical frequency metrology," Nature **416**, 233-237 (2002).

[2] Th. Udem, J. Reichert, R. Holzwarth, and T. W. Hänsch, "Absolute optical frequency measurement of the cesium $D_1$ line with a mode-locked laser," Phys. Rev. Lett. **82**, 3568 (1999).

[3] A. Cignöz, D. C. Yost, T. K. Allison, A. Ruehl, M. E. Fermann, I. Hartl, and J. Ye, "Direct frequency comb spectroscopy in the extreme ultraviolet," Nature **482**, 68-71 (2012).

[4] F. Keilmann, C. Gohle, and R. Holzwarth, "Time-domain mid-infrared frequency-comb spectrometer," Opt. Lett. **29**, 1542-1544 (2004).

[5] T. Yasui, Y. Kabetani, E. Saneyoshi, S. Yokoyama, and T. Araki, "Terahertz frequency comb by multifrequency-heterodyning photoconductive detection for high-accuracy, high-resolution terahertz spectroscopy," Appl. Phys. Lett. **88**, 241104 (2006).

[6] J. Mandon, G. Guelachvili, and N. Picqué, "Fourier transform spectroscopy with a laser frequency comb," Nature Photon, **3**, 99-102 (2009).

[7] B. Bernhardt, A. Ozawa, P. Jacquet, M. Jacqey, Y. Kobayashi, Th. Udem, R. Holzwarth, G. Guelachvili, T. W. Hänsch, and N. Picque, "Cavity-enhanced dual-comb spectroscopy," Nature Photon. **4**, 55-57 (2010).

[8] T. Ideguchi, A. Poisson, G. Guelachvili, T. W. Hänsch, and N. Picqué, "Adaptive dual-comb spectroscopy in the green region," Opt. Lett. **37**, 4847-4849 (2012).

[9] I. Coddington, W. C. Swann, and N. R. Newbury, "Time-domain spectroscopy of molecular free-induction decay in the infrared," Opt. Lett. **35**, 1395-1397 (2010).

[10] T. Yasui, M. Nose, A. Ihara, K. Kawamoto, S. Yokoyama, H. Inaba, K. Minoshima, and T. Araki, "Fiber-based, hybrid terahertz spectrometer using dual fiber combs," Opt. Lett. **35**, pp. 1689-1691 (2010).

[11] C. Janke, M. Först, M. Nagel, H. Kurz, and A. Bartels, "Asynchronous optical sampling for high-speed characterization of integrated resonant terahertz sensors," Opt. Lett. **30**, 1405-1407 (2005).

[12] T. Yasui, E. Saneyoshi and T. Araki, "Asynchronous optical sampling terahertz time-domain spectroscopy for ultrahigh spectral resolution and rapid data acquisition," Appl. Phys. Lett. **87**, 061101 (2005).

[13] T. Yasui, K. Kawamoto, Y.-D. Hsieh, Y. Sakaguchi, M. Jewariya, H. Inaba, K. Minoshima, F. Hindle, and T. Araki, "Enhancement of spectral resolution and accuracy in asynchronous-optical-sampling terahertz time-domain spectroscopy for low-pressure gas-phase analysis," Opt. Express 20, l15071–15078 (2012).

[14] P. A. Elzinga, R. J. Kneisler, F. E. Lytle, Y. Jiang, G. B. King, and N. M. Laurendeau, "Pump/probe method for fast analysis of visible spectral signatures utilizing asynchronous optical sampling," Appl. Opt. **26**, 4303-4309 (1987).

[15] Y.-D. Hsieh, Y. Iyonaga, Y. Sakaguchi, S. Yokoyama, H. Inaba, K. Minoshima, F. Hindle, Y. Takahashi, M. Yoshimura, Y. Mori, T. Araki, and T. Yasui, "Terahertz comb

spectroscopy traceable to microwave frequency standard," IEEE Trans. Terahertz Sci. Tech., in press (2013).

[16] T. Seta, H. Hoshina, Y. Kasai, I. Hosako, C. Otani, S. LoXowc, J. Urband, M. Ekström, P. Erikssond, and D. Murtaghd, "Pressure broadening coefficients of the water vapor lines at 556.936 and 752.033 GHz," J. Quantum Spectrosc. Radiat. Transfer **109**, 144–150 (2008).

[17] M. Kessler, H. Ring, R. Trambarulo, and W. Gordy, "Microwave spectra and molecular structures of methyl cyanide and methyl isocyanide," Phys. Rev. **79**, 54–56 (1950).

[18] D. M. Mittleman, R. H. Jacobsen, R. Neelamani, R. G. Baraniuk, and M. C. Nuss, "Gas sensing using terahertz time-domain spectroscopy," Appl. Phys. B **67**, 379–390 (1998).

[19] S. Matsuura, M. Tani, H. Abe, K. Sakai, H. Ozeki, and S. Saito, "High-resolution terahertz spectroscopy by a compact radiation source based on photomixing with diode lasers in a photoconductive antenna," J. Mol. Spectrosc. **187**, 97–101 (1998).

[20] H. M. Pickett, R. L. Poynter, E. A. Cohen, M. L. Delitsky, J. C. Pearson, and H. S. P. Muller, "Submillimeter, millimeter, and microwave spectral line catalog," J. Quant. Spectrosc. Radiat. Transf. **60**, 883–890 (1998).